\documentclass[11pt]{article}
\usepackage{amssymb,amsmath}
\usepackage{graphicx}
\newcommand{\ep}{\epsilon}

\newcommand{\vep}{\varepsilon}

\newcommand{\eps}{\varepsilon}

\newcommand{\be}{\begin{equation}}
\newcommand{\ee}{\end{equation}}
\newcommand{\bea}{\begin{eqnarray}}
\newcommand{\eea}{\end{eqnarray}}
\newcommand{\beas}{\begin{eqnarray*}}
\newcommand{\eeas}{\end{eqnarray*}}
\newcommand{\ba}{\begin{array}}
\newcommand{\ea}{\end{array}}

\newcommand{\tr}{\mathrm{Tr}}

\newcommand{\is}{\! &\! = \! & \!}
\newcommand{\iss}{= \! &\! }
\newpage

\renewcommand{\sp}{\hspace{1pt}}

\newcommand{\ccc}{\mbox{\large $ c$}}

\newcommand{\aaa}{\mbox{\large $ a$}}

\newcommand{\nbox}{{\,\lower0.9pt\vbox{\hrule \hbox{\vrule height 0.2 cm \hskip 0.19 cm \vrule height 0.2 cm}\hrule}\,}}

\def\href#1#2{#2}

\makeatletter
\@addtoreset{equation}{section}
\makeatother

\def\appendix#1{
  \addtocounter{section}{1}
  \setcounter{equation}{0}
  \renewcommand{\thesection}{\Alph{section}}
  \section*{Appendix \thesection\protect\indent \parbox[t]{11.15cm}
  {#1} }
  \addcontentsline{toc}{section}{Appendix \thesection\ \ \ #1}
  }

\newcommand{\deltaa}{\delta_{1}}
\newcommand{\deltas}{\delta_{}}

\begin{document}
\begin{titlepage}
\hfill
\vbox{
    \halign{#\hfil         \cr
           } % end of \halign
      }  % end of \vbox
\vspace*{15mm}
\begin{center}
{\Large \bf Supersymmetric Yang-Mills Theory  \\ From Lorentzian Three-Algebras}

\vspace{1.3truecm}
\centerline{
    {Jaume Gomis,${}^{a}$}\footnote{jgomis@perimeterinstitute.ca}
    {Diego Rodr\'{\i}guez-G\'omez,${}^{b,c}$}\footnote{drodrigu@princeton.edu}
   {Mark Van Raamsdonk,${}^{d}$}\footnote{mav@phas.ubc.ca}
    {Herman Verlinde,${}^{b}$}\footnote{verlinde@princeton.edu}}
   \vspace{1.4cm}
\centerline{{\it ${}^a$Perimeter Institute for Theoretical Physics}}
\centerline{{\it Waterloo, Ontario N2L 2Y5, Canada}}
\vspace{.5cm}
\centerline{{\it ${}^b$Department of Physics,
Princeton University}}
\centerline{{\it Princeton, NJ 08544, USA}}
\vspace{.5cm}
\centerline{{\it ${}^c$Center for Research in String Theory, Queen Mary University of
    London}} \centerline{{\it Mile End Road, London, E1 4NS, UK}}
\vspace{.5cm}
\centerline{{\it ${}^d$Department of Physics and Astronomy,
University of British Columbia}}
\centerline{{\it 6224 Agricultural Road,
Vancouver, B.C., V6T 1Z1, Canada}}

\vspace*{1cm}
%%\maketitle
\end{center}

\centerline{\bf ABSTRACT}
\vspace*{10mm}

We show that by adding a supersymmetric Faddeev-Popov ghost sector to the recently constructed Bagger-Lambert theory based on a Lorentzian three algebra, we obtain an action with a BRST symmetry that can be used to demonstrate the absence of negative norm states in the physical Hilbert space. We show that the combined theory, expanded about its trivial vacuum, is BRST equivalent to a trivial theory, while the theory with a vev for one of the scalars associated with a null direction in the three-algebra is equivalent to a reformulation of maximally supersymmetric 2+1 dimensional Yang-Mills theory in which there a formal $SO(8)$ superconformal invariance.

\end{titlepage}

\vskip 1cm

\addtolength{\baselineskip}{1.1mm}
\addtolength{\parskip}{.5mm}

\setcounter{equation}{0}
\section{Introduction}

The quest for a Lagrangian description of the worldvolume theory of coincident M2-branes in M-theory has been a problem of longstanding interest. Important guidance comes from space-time symmetries, most notably  $SO(8)$ invariance and maximal supersymmetry,  and from the relation between M2-branes in compactified M-theory and  D2-branes in Type  IIA string theory.
A further constructive insight is to sharpen the formulation of the problem via the identification of a specific decoupling limit
in which the M2-brane worldvolume  degrees of freedom dynamically decouple from the surrounding  eleven dimensional supergravity modes. Concretely, this decoupling is achieved by restricting the worldvolume
theory to  the very long wavelength modes.
In this limit, the M2-brane theory should take the form of a  2+1-dimensional local quantum field theory with $SO(8)$ superconformal symmetry.
Until very recently, however, it was not known how to write down any nontrivial theories of this sort, apart from a free theory describing a single M2-brane.

The work of Bagger and Lambert \cite{Bagger:2006sk}\cite{Bagger:2007jr}\cite{Bagger:2007vi}, and Gustavsson \cite{Gustavsson:2007vu}\cite{Gustavsson:2008dy}, following earlier work \cite{Basu:2004ed}\cite{Schwarz:2004yj}, showed that a manifestly supersymmetric and $SO(8)$ invariant Lagrangian can be constructed given a ``three-algebra," a generalization of a Lie algebra based on an antisymmetric triple product structure. It seemed plausible that the worldvolume theory of coincident M2-branes should lie somewhere in this class of theories. Initially, however, only a single example of a three-algebra was known, and this example has now been proven \cite{Papadopoulos:2008sk}\cite{Gauntlett:2008uf}
to be the only nontrivial example with positive definite metric (apart from direct sums of copies of this algebra with a trivial algebra). The corresponding field theory does appear to be related to M2-branes, in particular two M2-branes, but on an M-theory orbifold \cite{VanRaamsdonk:2008ft}\cite{Lambert:2008et}\cite{Distler:2008mk}.\footnote{The details of the orbifold depend on a discrete parameter that is present in the theory, the level associated with a Chern-Simons term.}

A proposal for the worldvolume theory of coincident M2-branes in flat space has been
made in \cite{Gomis:2008uv}\cite{Benvenuti:2008bt}\cite{Ho:2008ei}, henceforth called the BF membrane model. This theory is based on a Lorentzian ``three-algebra,"\footnote{The fact that the algebras in \cite{Gomis:2008uv}\cite{Benvenuti:2008bt}\cite{Ho:2008ei} are the unique undecomposable three-algebras has been proven in \cite{FigueroaO'Farrill:2008zm}.}
and its gauge structure is that of a three dimensional BF theory. The BF membrane model captures some of the expected properties of the theory on multiple M2-branes, including the absence of a coupling constant, a close relation \cite{Mukhi:2008ux} to three dimensional SYM on D2-branes and a moduli space corresponding to $N$ M2-branes in flat space.

As already mentioned in \cite{Gomis:2008uv}\cite{Benvenuti:2008bt}\cite{Ho:2008ei}, the BF membrane model has ghosts in the classical theory, arising from the timelike direction in the ``three-algebra". Despite the fact that ghosts  cannot propagate in loops due to the special nature of the interactions of the BF membrane model, it remains an important issue to understand whether the theory has negative norm states.

In this paper we enrich the BF membrane model by adding to the theory a set of supersymmetric Faddeev-Popov ghosts, preserving all the desired symmetries expected from the worldvolume theory on M2-branes. We show that the combined action is symmetric under a BRST transformation, suggesting that it should be interpreted as a gauge-fixed version of a classical action with an additional gauge symmetry. This acts as a gauged shift symmetry for bosonic and fermionic fields associated with the wrong-sign kinetic terms. Via a BRST analysis, we show that this extra symmetry is precisely what we need to ensure the absence of and negative norm states in the physical Hilbert space.

BRST-invariant actions arise as the gauge-fixed versions of classical actions with gauge-symmetries. It is therefore natural to ask which classical action our BRST-invariant action arises from. For the theory expanded about its trivial vacuum, we find that the classical action is a BF-theory coupled to matter without interactions. This theory has trivial dynamics. With a nonzero vev for $X^I_+$ (one of the scalars associated with the light-like directions in the three-algebra), we find that the classical theory is one that may be shown to be equivalent to 2+1 dimensional maximally supersymmetric Yang-Mills theory, though it does have a formal $SO(8)$ superconformal symmetry if we formally allow transformations of the $X^I_+$ vev. However, the theory does not directly describe the IR limit of the Yang-Mills theory, so it gives only the previously known indirect description of M2-branes in flat space.\footnote{In the original version of this paper, it was proposed that by integrating over the $X^I_+$ vev in the path integral, superconformal invariance could be restored. We no longer feel that the justification for doing so is correct.}

It may be that the new formulation offers advantages over the conventional description of the D2-brane theory. In support of this, we show that by dualizing the gauge field associated with the noncompact gauge symmetry of the theory, we are able to write down operators which are gauge invariant and BRST invariant, and  which  transform  in non-trivial representations of the formal $SO(8)$ R-symmetry. As an example, we write down a set of operators in the same representation of the superconformal algebra as the chiral primary operators for the M2-brane theory. Calculating correlation functions of these operators in the D2-brane theory in a limit where $X^I_+$ goes to infinity may give an algorithm for using the D2-brane theory to calculate arbitrary correlation functions in the M2-brane theory.

The plan of the paper is as follows. In section $2$ we review the main features of the BF membrane model constructed in \cite{Gomis:2008uv}\cite{Benvenuti:2008bt}\cite{Ho:2008ei}. In section $3$ we gauge the shift symmetry in \cite{Gomis:2008uv}\cite{Benvenuti:2008bt}\cite{Ho:2008ei} and gauge fix all the associated gauge symmetries and  show that the gauge fixed action is described by the Lagrangian in  \cite{Gomis:2008uv}\cite{Benvenuti:2008bt}\cite{Ho:2008ei} together with a supersymmetric  ghost action. By analyzing the
BRST transformations of the gauge fixed action we show that the theory has no negative norm states.
In section $4$ we study the gauge fixed action and make contact with the theory on multiple D2-branes. In section $5$ we discuss gauge invariant operators in our theory and show that we can construct operators in full multiplets of the formal $SO(8)$ representations. Section $6$ contains conclusions and some technical details are presented in the Appendices.

%\newpage

\setcounter{equation}{0}
\section{The BF membrane model}

Our starting point is the action derived in  \cite{Gomis:2008uv}\cite{Benvenuti:2008bt}\cite{Ho:2008ei}. These authors constructed a three-algebra with Lorentzian metric corresponding to an arbitrary Lie algebra $G$. The action for the corresponding Bagger-Lambert-Gustavsson theory is given by
\begin{eqnarray}
\cal{L}\is -\frac{1}{2}{\rm Tr}\Big((D_{\mu}X^I - B_\mu X^I_+)^2 \Big) + \partial_{\mu} X^I_+ (\partial_{\mu} X_-^I - {\rm Tr}(B_\mu X^I)) + \frac{i}{2}{\rm Tr}\Big(\bar{\Psi}\Gamma^{\mu}(D_{\mu}\Psi- B_\mu \Psi_+)\Big)\nonumber \\[1mm] &&- \frac{i}{2}\bar{\Psi}_+\Gamma^{\mu}(\partial_{\mu}\Psi_- - {\rm Tr}( B_\mu \Psi)) -\frac{i}{2}\bar{\Psi}_-\Gamma^{\mu}\partial_{\mu}\Psi_+ + \epsilon^{\mu\nu\lambda} {\rm Tr}\Big( B_{\lambda} (\partial_{\mu} A_{\nu} - [A_{\mu}, A_{\nu}]) \Big) \nonumber\\[1mm]
&& - \frac{1}{12} {\rm Tr}\Big(X_+^I [ X^J, X^K] + X^J_+ [ X^K , X^I ] + X_+^K [ X^I , X^J ]\Big)^2 \label{action}
\\[1mm]
&& +\frac{i}{2}{\rm Tr}\Big(\bar{\Psi}\Gamma_{IJ}X_ +^I[X^J,\Psi]\Big)+\frac{i}{4}{\rm Tr}\Big(\bar{\Psi}\Gamma_{IJ}[X^I,X^J]\Psi_+\Big)-\frac{i}{4}{\rm Tr}\Big(\bar{\Psi}_+\Gamma_{IJ}[X^I,X^J]\Psi\Big)\ .
\nonumber
\end{eqnarray}
Here $D_\nu = \partial_\nu -2 [A_\nu, \cdot\, ]$,
with $A_\mu$  a gauge field for the compact gauge group $G$.
The fields  $X^I$, $\Psi$, and $B_\mu$ transform in the adjoint representation for this gauge symmetry,
while the fields $X^I_+$,$X^I_-$,$\Psi_+$, and $\Psi_-$ are singlets. The above
action does not contain a standard Yang-Mills kinetic term for the gauge boson $A_\mu$, but rather
a term of the  $B \wedge F$ form, which underlies the symmetry structure of  (\ref{action}). Via the presence of this additional one-form field $B$,
the theory has an extra non-compact gauge symmetry, under which the fields transform infinitesimally as
\bea
\label{Bgauge}
\deltaa B_\mu \is D_\mu \zeta\, ; \qquad \qquad
\deltaa X^I = \, \zeta X^I_+ \, ; \qquad \qquad
\deltaa X^I_-  =\, \tr(\zeta X^I) \, ; \nonumber\\[1mm]
\deltaa \Psi \is \zeta \Psi_+ \, ; \qquad \qquad \deltaa \Psi_- = \tr(\zeta \Psi) \, .
\eea
The non-compact and compact symmetry  together combine into a gauge invariance under the
(2 dim $G$)-dimensional gauge group given by the Inonu-Wigner contraction of $G \otimes G$, which corresponds to the semidirect sum of the of translation algebra with the Lie algebra $G$, where $G$ acts on the $\hbox{dim} \; G$ translation generators in the obvious way.

The theory described by the above Lagrangian has several non-trivial properties, that support its interpretation as the multi-M2
brane worldvolume theory.

\medskip

\noindent
(i) Most remarkably, the Lagrangian (\ref{action}) is  invariant under a full $SO(8)$ superconformal symmetry.
The supersymmetry transformations read
\bea
\label{susy}
\deltas X^I\!  \is i \bar{\epsilon} \Gamma^I \Psi ;  \qquad \qquad
\deltas X^I_+  = \, i \bar{\epsilon} \Gamma^I \Psi_+ \, ;\qquad\qquad
\deltas X^I_- \! =\,  i \bar{\epsilon} \Gamma^I \Psi_-  \, ; \nonumber \\[2.2mm]
\deltas A_\mu  \is  {i \over 2} X^I_+ \bar{\epsilon} \Gamma_\mu \Gamma_I\Psi - {i\over 2} X^I \bar{\epsilon} \Gamma_\mu \Gamma^I \Psi_+ \, ;
 \ \ \quad \qquad \qquad
\deltas B_\mu  = {i} \bar{\epsilon} \Gamma_\mu \Gamma^I [X^I, \Psi] \, .\nonumber\\[1.8mm]
\deltas \sp \Psi\; \is (D_\mu X^I -B_{\mu}X_+^I)\, \Gamma^\mu \Gamma^I \epsilon - {1 \over 2} X^I_+ [X^J, X^K] \Gamma^{IJK} \epsilon\,  ;  \quad \qquad \,
\deltas \Psi_+ =\,   \partial_\mu X^I_+ \Gamma^\mu \Gamma^I \epsilon\, ; \nonumber \\[1mm]
\deltas \Psi_- \! \is (\partial_\mu X^I_- - \tr(B_\mu X^I_-)) \Gamma^\mu \Gamma^I \epsilon -{1 \over 3} \tr(X^I X^J X^K) \Gamma^{IJK} \epsilon \, ;
\eea

\smallskip

\noindent
(ii) The theory described by (\ref{action}) does not have a tunable coupling constant: overall multiplication of the Lagrangian by a constant can be absorbed into a suitable rescaling of the
fields. $X^I \to g\, X^I$,  $X^I_+ \to g^{-1}\,X_+^I$ $X_-^I\to g^3\, X_-^I$, ${B} \to
g^2\,B $.

\smallskip

\noindent
(iii) In case ${G} = U(N)$, the moduli space of vacua contains a branch of the form ${(\mathbb{R}^8)^{N}}/{S_N}$, which  is  as expected for a theory describing $N$ $M2$ branes in flat eleven dimensional space. However,  it is the $Osp(8|4)$ invariant vacuum of the theory in the unbroken phase  which  holographically describes   M-theory in $AdS_4 \times S^7$. The states of the BF model in the broken phase correspond to half supersymmetric geometries which are  $AdS_4 \times S^7$ only asymptotically.

\smallskip

\noindent
(iv) When restricted to (or, as we shall see, expanded around) the background with constant $X_+$ and $\Psi_+=0$, the Lagrangian (\ref{action}) reduces to that of 2+1-dimensional SYM theory, the low energy theory on $N$ D2-branes, where the SYM coupling constant is identified with $g^2_{YM}=X_+^2$ \cite{Ho:2008ei}.

\medskip

\noindent
Given all these properties, the theory described by (\ref{action}) appears to be a step in the right direction towards finding the multi-M2 brane worldvolume theory.

An outstanding problem, however, is that the theory (\ref{action})  includes fields, $X_+$ and $X_-$ with a non-positive definite kinetic term, and which arise from the necessity of using a three-algebra with Lorentzian signature. We now proceed to enrich the theory in (\ref{action}) and show that the enriched model has no negative norm states.

\section{Eliminating the ghosts -- by adding ghosts}

An unsettling feature of the theory (\ref{action}), arising from the non-positivity of the three-algebra metric, is that a set of scalar fields\footnote{Where we define $A_\pm=A_0\pm A_1$.} $X^I_0$ and fermionic fields $\Psi_0$
have a wrong-sign kinetic term. It has been suggested that this may not pose a problem if we make a projection on the space of physical states, or simply because the $X^I_+$ and $X^I_-$ fields cannot propagate in loops (since the interaction terms depend on $X^I_+$ but not $X^I_-$).

We   argue that by gauging the global shift symmetry
\bea
X^I_- \to X^I_- + a^I\cr
\Psi_-\to \Psi_- + \chi
\eea
of the action (\ref{action}) and properly gauge fixing all the associated gauge symmetries, that the gauged fixed Lagrangian is described by the original theory (\ref{action}) together with a  supersymmetric Faddeev-Popov ghost action.
Moreover, we show that the gauge fixed theory is free of negative norm states.

\subsection{Gauging the shift symmetry}

We start by introducing bosonic and fermionic gauge fields $a^I_\mu$ and $\eta_\mu$. These gauge fields  are
  associated with gauging the shift symmetry of  $X_-$ and its superpartner $\Psi_-$
\bea
\delta X^I_- \is \beta^I , \qquad \qquad  \quad\delta \Psi_- = \chi \,  ;
\nonumber \\[1.5mm]
%\qquad \quad
\delta a^I_\mu \is \partial_\mu \beta^I ,  \qquad \qquad \delta \eta_\mu\, = \partial_\mu \chi\, .
\eea
The Lagrangian (\ref{action}) can be made invariant under these new gauge transformations, if we add the terms
\be
\label{terms}
\tilde{\cal L} = -a^{I}_\mu \partial^\mu X^I_+ - i  \bar{\eta}_\mu \Gamma^\mu \Psi_+ \;,
\ee
which turn the $X^I_-$ and $\Psi_-$ derivatives into covariant derivatives. It can be shown  that the modification of the action preserves supersymmetry if we supplement the supersymmetry variations (\ref{susy}) with the additional terms
\beas
\label{newsusy}
\tilde{\delta} \Psi_- &=& - \Gamma^\mu \Gamma^I \epsilon \; a_\mu^I \cr
\tilde{\delta} a_\mu^I &=& i \bar{\epsilon} \Gamma^I \Gamma_\mu \Gamma^\nu \eta_\nu \cr
\tilde{\delta} \eta_\mu &=& -{1 \over 2} (\partial_\mu a^I_\nu - \partial_\nu a^I_\mu) \Gamma^\nu \Gamma^I \epsilon\;.
\eeas
Note that these   supersymmetry variations commute with the shift gauge symmetry and that  (\ref{newsusy}) generate the expected supersymmetry algebra on the fields, as we show in  Appendix $A$.\footnote{The commutator of two supersymmetry transformations generate on-shell   a translation and gauge transformations.}

The presence of an extra gauge symmetry here appears to be just what we need to save the theory from negative norm states. By choosing a gauge $X^I_-=0$, and $\Psi_-=0$, we completely eliminate the problematic kinetic term, and the remaining integrals over $c^I_\mu$ and $\eta_\mu$ appear to force $\partial_\mu X^I_+ =0$ and $\Psi_+=0$. However, the new Lagrangian has extra gauge symmetries, arising from the fact that only $\partial_\mu a^{I \mu}$ and $\gamma^\mu \eta_\mu$ appear in the action, so care must be taken in gauge fixing these new symmetries. In detail (\ref{terms})  is invariant under\footnote{Note that $\alpha^{\rho I}$ and $\chi^\nu$ do not all correspond to independent gauge transformations since the gauge transformations corresponding to $\alpha^{\rho I} = \partial^\rho \beta^I$ and $\chi^\nu = \Gamma^\nu \xi$ are trivial. They correspond to {\it reducible} gauge transformations.}
\be
a_\mu^I \to a_\mu^I + \eps_{\mu\nu\rho} \partial^\nu \alpha^{\rho I}
\ee
and
\be
\label{larger}
\eta_\mu \to \eta_\mu + \chi_\mu - {1 \over 3} \Gamma_\mu \Gamma_\nu \chi^\nu \; .
\ee
Because of this extra gauge symmetry, simply integrating over $c^I_\mu$ and $\eta_\mu$ after fixing the gauge $X^I_- = \Psi_-=0$   leads to a divergence in the path integral.\footnote{The set of delta functions $\delta(\partial_\mu X^I_+)$ that we end up with are too many to leave the single integral over $X^I_+$ finite.}

A careful discussion of the quantization of this theory taking into account the additional gauge symmetries is presented in Appendix $B$. The result of this analysis is that the gauge fixed action is described by the Lagrangian (\ref{action})  with the addition of a supersymmetric ghost action, which we now analyze in some detail.

\subsection{Interpretation as a gauge-fixed action}

In the previous section, we have seen that it is possible to promote the $X_-$ and $\Psi_-$ shift symmetries to gauge symmetries in a way that is consistent with superconformal invariance. Taking for granted that this symmetry should be understood as a gauge symmetry and doing the proper gauge fixing procedure, as described in Appendix $B$, we find that the gauge fixed theory can interpreted as  the  original action (\ref{action})   with a corresponding Faddeev-Popov ghost action
\begin{eqnarray}\label{LAGa}
\mathcal{L}_{\rm ghost} &=&-  \partial^{\mu}\ccc^I_- \partial_{\mu}\ccc^I_+ + i  \bar{\chi}_+\Gamma^{\mu}\partial_{\mu}\chi_-\;,
\end{eqnarray}
where $\ccc^I_\pm$ are 8 anti-commuting scalars, and $\chi_\pm$ are commuting SO(8) spinors. This ghost action is ${\cal N} =8$ supersymmetric, and is invariant under the following supersymmetry transformations
\begin{equation}
\label{ghostsusy}
\delta \ccc^I_\pm=i\bar{\epsilon}\sp \Gamma^I\chi_\pm\ ,\qquad \qquad \delta\chi_\pm=\partial_{\mu}\ccc^I_\pm \Gamma^{\mu}\Gamma_I\sp \epsilon\, .
%&& \delta (\tilde{A}_{\mu})^a_b=i\bar{\epsilon}\Gamma_{\mu} \Gamma_I X^I_c \Psi_d f^{cda}\,_b\ ;\nonumber
\end{equation}

The action (\ref{action}) combined with the ghost action (\ref{LAGa}) is invariant under the nilpotent BRST transformations
\begin{eqnarray}
\label{brst}
\delta_{\rm brst} X_-^I =  \varepsilon \sp \ccc^I_-\, ,  & \qquad & \delta_{\rm brst} \Psi_- = \varepsilon\sp \chi_- \nonumber \\[2mm]
\delta_{\rm brst} \ccc_+^I =  \varepsilon\sp {X}_+^I\, ,  & \qquad & \delta_{\rm brst} \chi_+ = \varepsilon\sp \Psi_+.
\end{eqnarray}
The fact that the original action combined with the ghost action has a BRST symmetry implies that we should think of the combination as a gauge-fixed action. As usual, the corresponding gauge symmetry of the `unfixed' theory is given by the BRST transformation of the matter fields, so we see that it is precisely the shift gauge symmetry for the fields $X^I_-$ and $\Psi_-$.

\subsection{Physical states}

Physical states are defined by the BRST cohomology:
\begin{eqnarray}
Q_{\rm brst} |{\rm phys} \rangle =  0 \, , \qquad \qquad
|{\rm phys} \rangle  \equiv  |{\rm phys} \rangle + Q_{\rm brst} | {\rm anything}\rangle\, .
\end{eqnarray}
The BRST invariance condition can be solved trivially by requiring that all positive frequency (= annihilation) modes
of $\ccc_-^I$, $X_+$, $\chi_-$ and $\Psi_+$ annihilate the physical states. All states created
by negative frequency (= creation) modes of the same four fields are spurious: they are orthogonal
to this particular subspace, and in fact, can be written as $Q_{\rm brst} | {\rm something}\rangle$.

\subsection{A simple example: the $U(1)$ theory}

To see explicitly that our BRST-invariant action results in only positive-norm states, let us consider the simplest nontrivial example, the BF theory for gauge group $U(1)$.

We will see that this theory is precisely equivalent to the theory of single membrane, the free superconformal theory
\[
{\cal L} = -{1 \over 2} \partial_\mu X^I \partial^\mu X^I + {i \over 2} \bar{\Psi} \Gamma^\mu \partial_\mu \Psi \; .
\]
Now, the BF theory in the $U(1)$ case reduces to
\beas
{\cal L} &=& -{1 \over 2} (\partial_\mu X^I- B_\mu X^I_+)^2 + \partial_\mu X^I_+ \partial^\mu X^I_- - \partial_\mu X^I_+ B_\mu X^I + {1 \over 2} \epsilon^{\mu \nu \lambda} B_\mu F_{\nu \lambda} \\[1.5mm]
&& + {i \over 2} \bar{\Psi} \Gamma^\mu \partial_\mu \Psi -i \bar{\Psi}_+ \Gamma^\mu (\partial_\mu \Psi_- - B_\mu \Psi) + {\cal L}_{ghost}.
\eeas
Here, the ghost action ${\cal L}_{ghost}$ is given in (\ref{LAGa}) and  $F_{\mu \nu}$ is the field strength for $A_\mu$. We can treat this field strength as an independent variable if we introduce a Lagrange multiplier term
\[
\sigma \epsilon^{\mu \nu \lambda} \partial_\mu F_{\nu \lambda},
\]
to enforce the Bianchi identity. Here $\sigma$ is the dual gauge scalar, that, after integrating out $F_{\nu\lambda}$, furnishes a `magnetic' dual description of the $A_\mu$ degrees of freedom.\footnote{Note
that  in the vacuum with $X_+^I = v^I$, the equation
of motion of $B_\mu$ identifies $B_\mu = \frac{1}{v^2}\epsilon_{\mu\nu\lambda} F^{\nu\lambda}$.}
The equations of motion for $F_{\mu \nu}$ then give
\be
\label{dualboson}
B_\mu  =  \partial_\mu \sigma \; .
\ee
Typically, it is argued that the dual gauge scalar field $\sigma$ is naturally defined to be periodic.
Since $B$ is associated with a non-compact gauge symmetry, the equation (\ref{dualboson}) indicates that $\sigma$ should be allowed to range over the full real axis. Making the substitution for $B$ here gives
\beas
{\cal L} &=& -{1 \over 2} (\partial_\mu X^I-  \partial_\mu \sigma X^I_+)^2 + \partial_\mu X^I_+ \partial^\mu X^I_- -  \partial_\mu X^I_+  \partial_\mu \sigma X^I \\[1.5mm]
&=&+ {i \over 2} \bar{\Psi} \Gamma^\mu \partial_\mu \Psi - i \bar{\Psi}_+ \Gamma^\mu (\partial_\mu \Psi_- - \partial_\mu \sigma \Psi) + {\cal L}_{ghost}
\eeas
The field $\sigma$ transforms under the $B$ gauge symmetry as
\[
\sigma \to \sigma + \zeta
\]
so we can fix this gauge symmetry by setting $\sigma$ to zero. The resulting action is simply
\bea
{\cal L} &=& -{1 \over 2} \partial_\mu X^I \partial^\mu X^I  + {i \over 2} \bar{\Psi} \Gamma^\mu \partial_\mu \Psi \nonumber  \\[1.5mm]
&&+ \partial_\mu X^I_+ \partial^\mu X^I_- - i \bar{\Psi}_+ \Gamma^\mu \partial_\mu \Psi_-  -  \partial^{\mu}\ccc^I_- \partial_{\mu}\ccc^I_+ + i  \bar{\chi}_+\Gamma^{\mu}\partial_{\mu}\chi_-\; , \nonumber
\eea
so we end up with the free superconformal theory plus a free BRST invariant and superconformal invariant action for the $+/-$ fields and the ghosts.

The   BRST transformations (\ref{brst}) imply that acting with creation operators corresponding to $X^I_-$, $c^I_+$, $\Psi_-$, or $\chi_+$ on any physical state takes us out of the physical subspace, while acting with creation operators corresponding to $X^I_+$, $c^I_-$, $\Psi_+$, or $\chi_-$ on any physical state gives us BRST exact states. Thus, the BRST cohomology for the theory in the second line is trivial, so the BRST-invariant $U(1)$ BF theory is precisely equivalent to the free superconformal theory describing a single membrane.

\section{Classical Action}

It is natural to ask whether the BRST-invariant action we have found is the gauge-fixed version of some classical action. We now consider this question.

\subsection{The theory for constant $X^I_+$}

To start, consider the original action (\ref{action}), taking $X^I_+ = v^I$ constant and setting $\Psi_+ = 0$. The resulting action is
\bea
{\cal L}_0&=&-\frac{1}{2}{\rm Tr}\Big((D_{\mu}X^I - B_\mu v^I)^2 \Big) + \frac{i}{2}{\rm Tr}\Big(\bar{\Psi}\Gamma^{\mu}D_{\mu}\Psi \Big)  + {1 \over 2} \epsilon^{\mu\nu\lambda} {\rm Tr}\Big( B_{\lambda} F_{\mu \nu} \Big) \label{Lzero}
\\[1.5mm] && - \frac{1}{12} {\rm Tr}\Big(v^I [ X^J, X^K] + v^J [ X^K , X^I ] + v^K [ X^I , X^J ]\Big)^2 +\frac{i}{2}{\rm Tr}\Big(\bar{\Psi}\Gamma_{IJ}v^I[X^J,\Psi]\Big) .\nonumber
\eea
Note that the constant value of $X^I_+$ breaks conformal invariance and breaks $SO(8)$ invariance to $SO(7)$ invariance. Remarkably, as pointed out by \cite{Ho:2008ei}, this action for any nonzero $v^I$ turns out to be exactly the low-energy D2-brane action, i.e. maximally supersymmetric 2+1 dimensional Yang-Mills theory. For example, in the case where $v^I = g \delta^{I 8}$, we note that the $B$ gauge symmetry can be used to set $X^8=0$ , while the remaining integral over $B$ gives
\beas
{\cal L} &=& \tr\left(-{1 \over 4 g^2}F_{\mu \nu} F^{\mu \nu}-{1 \over 2} (D_\mu X^i)^2 + {i \over 2} \bar{\Psi} \Gamma^\mu D_\mu \Psi + {i \over 2} g \bar{\Psi}\Gamma^8\Gamma_{i}[X^i,\Psi] + {g^2 \over 4} [X^i,X^j]^2\right)
\eeas
where $i=1, \dots,7$.

\subsection{Reduction of the full theory to constant $X^I_+$}

Now, in our BRST-invariant action, we do not want to take $X^I_+$ or $\Psi_+$ to be constant. However, consider now the theory with some nonzero vev $X^I_+ = v^I$ where $X^I_+$ and all other fields are taken to be dynamical. Using the BRST transformations above, it is not hard to see that the complete action takes the form
\[
{\cal L}_0 + \{Q, \Xi \} \; .
\]
That is, our full action differs from the simple action (\ref{Lzero}) by a term that is BRST exact.\footnote{It is important to recall that the zero mode of $X^I_+$ effectively does not appear in the BRST transformation otherwise, we would conclude that even the $v^I$-dependent terms in ${\cal L}_0$ were BRST exact.} In fact, any BRST invariant action may be written in this way, and one typically identifies ${\cal L}_0$ with the classical action and the BRST-exact term as a gauge-fixing term.\footnote{Note that while $X^I_-$ and $\Psi_-$ do not appear in the classical action, we should consider them to be part of the classical fields, since their shift-gauge symmetry is the one we are fixing by going to the BRST-invariant action. On the other hand, the $X^I_+$ and $\Psi_+$ fields play the role of `$h$' fields (or `Nakanishi-Lautrup' fields) in the BRST-invariant action, and should not be thought of fields in the classical action.} Different choices of $\Xi$ do not change the physical observables (though one typically chooses $\Xi$ so that all global symmetries of the theory are preserved). Thus, it appears that for a fixed non-zero value of the zero mode of $X^I_+$, our theory simply corresponds to a particular gauge-fixing of the D2-brane theory plus $X_-$ and $\Psi_-$ fields that do not appear in the classical Lagrangian.

A special case in which we do preserve superconformal invariance is to set $X^I_+=0$. However, in this case we reach
the more severe conclusion that the full action can be written as a BRST exact piece plus the simple action
\be
\label{classical}
{\cal L}_0 = {\rm Tr}\Big(-\frac{1}{2}(D_{\mu}X^I )^2 + \frac{i}{2}
\bar{\Psi}\Gamma^{\mu}D_{\mu}\Psi + {1 \over 2} \epsilon^{\mu\nu\lambda}
B_{\mu} F_{\nu \lambda} \Big)
\ee
Note that this action by itself is superconformally invariant and
invariant under the $B$ gauge transformation (we simply set all the terms
in the transformation laws involving the + and - fields to zero). However,
at least naively, the theory appears to be free, giving $N^2$ copies of
the free superconformal multiplet, since integrating out the $B$ gauge
field seems to eliminate all the degrees of freedom of $A$. This is also
consistent with the point that for finite $v^I$ we obtain a D2-brane
theory with coupling $g^2 =v^2$: for length scales $L \ll 1/v^2$
correlation functions are trivial, and this length scale goes to infinity
in the limit.

\subsection{Summary}

In the end, it is reasonable to ask whether we have really gained anything here, since we have ended up with either a trivial superconformal theory or a reformulation of the D2-brane theory, where one expects to recover the full SO(8) superconformal invariance only at the infrared fixed point \cite{Banks:1996my}, \cite{Sethi:1997sw}. However, the formal $SO(8)$ invariance that we have restored may be useful, and as we will emphasize in the next section, one important advantage with the new description is that we can write down the full $SO(8)$ multiplets of physical operators.

\section{Observables}

The basic physical observables for our theory are correlation functions of operators which are BRST-invariant and also gauge-invariant under the remaining non-abelian gauge symmetries.\footnote{In principle, we could also consider correlation functions of BRST-invariant operators which are not gauge-invariant, but whose gauge-variation is BRST exact. The correlation function of such operators will be gauge-invariant, since the physical states are BRST invariant.}
As we mentioned earlier, one of the nice features this formalism is that we will be able to construct full $SO(8)$ multiplets of operators.

\subsection{$SO(8)$ invariant basis}

Note first that operators such as
\[
{\cal O}^{I_1 \cdots I_n} = \tr(X^{I_1} \cdots X^{I_n}),
\]
while invariant under the compact gauge symmetry, are not invariant under the noncompact gauge transformations associated with the $B_\mu$ gauge field
\[
\delta B_\mu = D_\mu \zeta \qquad \qquad \delta X^I = X^I_+ \zeta \; .
\]
Hence it appears that the longitudinal component of $X^I$ (the one parallel to $X_+^I$) is spurious,
and that only the 7 transverse components of $X^I$ (the ones orthogonal to $X_+^I$) survive as
physical degrees of freedom. The scalar fields $X^I$ by themselves do not appear to be sufficient to
build a full $SO(8)$ multiplet of physical observables.

A useful hint for how to obtain these operators is provided via the relation of our theory to 2+1 SYM. The SYM theory only has $SO(7)$ symmetry. $SO(8)$ invariance emerges via a specific reordering of the degrees of freedom in the IR theory, whereby the $A_\mu$ gauge
boson gives rise to a dual scalar mode $\phi$, that combines with the other 7 scalar fields
into an SO(8) covariant vector. Thus we should expect that also in our theory, the full
$SO(8)$ multiplets must somehow involve the dual gauge field, and we will now see that this turns out to be correct.

It is possible to construct an $SO(8)$ covariant version of $X^I$ that does not transform under the $B$ gauge symmetry. To do this, we can consider the combination
\be
\label{covX}
Y^I = X^I - X^I_+ \phi
\ee
where $\phi$ is a field built from $A$ and $B$ with the noncompact gauge transformation law
\be
\label{phitf}
\delta \phi = \zeta\, .
\ee
Formally, we can take such a field to be determined in terms of $A$ and $B$ by the covariant equation
\be
\label{phidef}
D^2 \phi = D_\mu B^\mu \; ,
\ee
which is solved by
\[
\phi = {1 \over D^2} D_\mu B^\mu \; .
\]
It seems that $\phi$ is not really a local operator. However, in a gauge where $\phi=0$, the expression (\ref{covX}) reduces to $X^I$, so the construction appears to be local at least in this gauge choice.

We can introduce the scalar $\phi$ as a new independent field in the Lagrangian, by adding the term
\be
\label{gauge}
%\partial^\nu A_\nu  = 0 \, ; \qquad \quad
D^\nu \lambda( B_\nu - D_\nu \phi) + D^\nu \bar{\omega} D_\nu \omega \, .
\ee
Here $\lambda$ can be viewed as a dual Lagrange multiplier field, whose equation of motion
implies the identification (\ref{phidef}). The second term in (\ref{gauge}) is a ghost action, that ensures that the functional determinant produced by integrating out $\lambda$, $\phi$  cancels out.
The total action has the non-compact gauge invariance
\be
\label{residual}
%\delta A_\nu = D_\nu \alpha \; ; \qquad \quad
\delta B_\nu = %[\alpha, B_\nu] +
D_\nu \beta\, , \qquad
 \qquad
%\partial^\nu D_\nu \alpha = 0\, ; \qquad \quad
\delta \phi = \beta\, , \qquad \qquad \delta X^I = \beta X_+^I\, .
%+  [\alpha, B_\nu]) = 0\,
\ee
%where $D_\nu = \partial_\nu + [A_\nu, \cdot\, ]$.

But now let's use this invariance to choose the gauge fixing condition $\phi=0$. The action (\ref{gauge}) then turns into the standard gauge fixing
action, imposing the gauge condition
\be
D^\nu B_\nu = 0.
\ee
In this gauge, the operators (\ref{covX}) simply reduce to $Y^I = X^I$.

Geometrically, we can clarify the situation as follows.
The scalar field $\phi$ defined via (\ref{phidef}) can be recognized as the dual gauge scalar, the
``magnetic"  dual to the non-abelian gauge field $A$. Eqn (\ref{phidef}) should be compared
with  (\ref{dualboson}), that defines the dual gauge scalar for the abelian theory. As emphasized
earlier, this dual gauge boson is not periodic, but takes values on the full real axis.
The non-compact  gauge symmetry thus acquires an interesting geometric significance,
reflecting the physical equivalence between the dual gauge boson and the longitudinal component
of the scalar fields $X_I$, or rather, that the two scalar fields in fact constitute only one single
scalar degree of freedom. The gauge choice $\phi = 0$ represents a physical gauge,
in which this dual gauge scalar degree of freedom is completely absorbed into the longitudinal
component of $X^I$.
%\footnote{
%In the vacuum with $X_+^I = v^I$, the equation
%of motion of $B_\mu$ identifies $B_\mu = \frac{1}{v^2}\epsilon_{\mu\nu\lambda} F^{\nu\lambda}$.}
%, so that$D^2 \phi = \frac{1}{v^2}\epsilon_{\mu\nu\lambda} D_\mu F^{\nu\lambda}$

A priori, using the $\phi=0$ gauge, we can just use the fields $X^I$ to build physical observables.
There is a slight subtlety, however.
The gauge fixed theory still has a BRST invariance
\be
\delta_1 B_\nu = %[\alpha, B_\nu] +
\varepsilon D_\nu \omega\, , \qquad
 \qquad
%\partial^\nu D_\nu \alpha = 0\, ; \qquad \quad
\delta_1 X^I = \varepsilon\sp \omega X_+^I \, ,  \qquad \qquad \delta_1 \bar \omega = \varepsilon \lambda\, ,
\ee
which explicitly does not leave $X^I$ invariant.
This invariance can be restored as follows.
Recall that the
M2-brane theory has another set of ghost fields, associated with the $X_-$ shift symmetry,
an therefore another BRST invariance, given in (\ref{brst}). Let us denote this BRST variation by
$\delta_2$:
\begin{eqnarray}
\delta_{2} X_-^I =  \varepsilon \sp \ccc^I\, ,  & \qquad &
% \delta_{2} \Psi_- = \varepsilon\sp \chi_- \nonumber \\[2mm]
\delta_{2} \ccc_+^I =  \varepsilon\sp {X}_+^I\, . %& \qquad & \delta_{2} \chi_+ = \varepsilon\sp \Psi_+
\end{eqnarray}
The combined BRST symmetry is the sum of the two transformations
\be
\label{sum}
\delta_{\rm brst} = \delta_1+ \delta_2.
\ee
Physical operators do not need to be invariant under both transformations $\delta_1$ and $\delta_2$
separately, but only under the sum.
The following operators
\be
Y^I = X^I + \omega\sp \ccc_+^I
\ee
are invariant under this combined symmetry. Physical observables are obtained by taking combinations
\be
{\cal O}^{I_1I_2 \ldots I_p} = {\rm Tr}(Y^{I_1} Y^{I_2} \ldots Y^{I_p} ),
\ee
and are invariant under the compact gauge symmetries and the BRST transformation (\ref{sum}).

For most types of correlations functions, the ghost terms in $Y^I$ are not expected to contribute.
So in practice, we can think of the $Y^I$ operators as simply being equal to $X^I$. Indeed,
one can argue that we for most practical purposes can use the physical operators
\[
{\cal O}^{I_1I_2 \ldots I_p} = {\rm Tr}(X^{I_1} X^{I_2} \ldots X^{I_p} )\, .
\]
These operators transform non-trivially under the non-compact gauge transformations.
However, via a similar argument as above, one can show that its gauge variation can be written
as a total variation under the $\delta_2$ BRST symmetry, and thus decouples from physical
correlation functions.

\medskip

\subsection{Chiral Primary Operators}

In the worldvolume theory of M2-branes, by the AdS/CFT correspondence, the local operators of finite dimension should be in one-to-one correspondence with the states of M-theory on $AdS^4 \times S^7$ \cite{Minwalla:1998rp}. In particular, in the large $N$ limit, we should have protected operators in one-to-one correspondence with the spectrum of supergravity fluctuations on $AdS^4 \times S^7$. These operators appear in infinite dimensional multiplets of the superconformal algebra, which may be generated by the action of superconformal generators on chiral primary operators. An important set of observables for the theory are the correlation functions of these operators and their descendants. In principle, one should be able to compute these correlation functions using correlation functions in the D2-brane theory by scaling all distances to infinity at the end. However, in the usual formulation, it is not clear how to write the required full multiplets of operators. 

In the reformulation of the D2-brane theory that we arrive at, we can write operators in full multiplets of the formal $SO(8)$ symmetry. In particular, operators in the same multiplets as the chiral primaries may be written as 
\[
\tr(Y^{(I_1} \cdots Y^{I_n)}) - {\rm traces}
\]
and the multi-trace generalizations, where $Y^I$ is defined in the previous section. Assuming that the formal $SO(8)$ symmetry that we find becomes the full quantum $SO(8)$ symmetry in the IR limit of the theory, it seems plausible that calculating correlation functions of these operators in our theory then taking the lengths scales to infinity (perhaps with a suitable rescaling of the operators) could give a way to correlation functions of chiral primary operators in the M2-brane theory starting from the D2-brane theory.

\bigskip

\section{Conclusions}

In this paper we have shown that if we interpret the BF membrane model of \cite{Gomis:2008uv}\cite{Benvenuti:2008bt}\cite{Ho:2008ei} as a gauged fixed action, and add to it a corresponding set of supersymmetric Faddeev-Popov ghosts, that the resulting theory is devoid of negative norm states. We have shown that for a constant  maximally supersymmetric SYM living on D2-branes.

In the process of studying local gauge invariant operators, we have found that the gauge field $B_\mu$ associated to the non-compact gauge symmetry of the BF membrane model can be dualized exactly into a scalar. This construction allows us to write down  gauge invariant and BRST invariant operators transforming in non-trivial  representations of the formal $SO(8)$ symmetry. This symmetry remains very obscure in the definition of the theory as the infrared limit of  three dimensional maximally supersymmetric SYM, where only an $SO(7)$ subgroup of the symmetry is manifestly realized. The action considered in this paper allows us to bypass these difficulties and give a direct $SO(8)$ covariant realization of operators. 

To conclude, we note that while the present treatment of our theory by adding Fadeev-Popov ghosts and interpreting the full action as a BRST-invariant gauge-fixed action does not lead directly to a non-trivial superconformally invariant quantum theory, it might be that some other treatment gives a more direct relation to M2-branes. 

\vspace*{10mm}

\noindent {\bf Note added:} While this paper was being finished, the paper
\cite{Bandres:2008kj}  appeared in the arXiv, which has some overlap  with this paper.

%%%%%%%%%%%%%%%%%%%%%%%%%%%%%%%%%%%%%%%%%%%%%%%%%
\section*{Acknowledgments}
%%%%%%%%%%%%%%%%%%%%%%%%%%%%%%%%%%%%%%%%%%%%%%%%%%
We would like to thank Sergio Benvenuti, Bobby Ezhuthachan, Seba Franco, Joaquim Gomis, Filippo Passerini, Costis Papageorgakis, Arkady Tseytlin, and particularly Jorge Russo for helpful comments and discussion.
Research at Perimeter Institute is supported by the Government
of Canada through Industry Canada and by the Province of Ontario through
the Ministry of Research and Innovation. J.G.  also acknowledges further  support by an NSERC Discovery Grant. The work of MVR is supported in part by the Natural Sciences and Engineering Research Council of Canada, by and Alfred P. Sloan Foundation Fellowship, and by the Canada Research Chairs Programme.D. R-G. acknowledges financial support from the European Commission through Marie Curie OIF grant contract no. MOIF-CT-2006-38381. The work of H.V. is supported in part by the National Science Foundation
under grant PHY-0243680.

%% A P P E N D I X  %%%%%%%%%%%%%%%

\setcounter{section}{0}

\appendix{ Supersymmetry algebra}

Here we would like to show that
\be
[\delta_1,\delta_2] \eta_\mu = v^\nu (\partial_\nu \eta_\mu -\partial_\mu\eta_\nu )\ ,
\qquad
[\delta_1,\delta_2]c_\mu^I = v^\nu (\partial_\nu c^I_\mu -\partial_\mu c^I_\nu )\ ,
\ee
with
\be
v^\mu =-2i\bar   \ep_2 \Gamma^\mu \ep_1
\ee
modulo  gauge transformations and equations of motion.

We will use the relation
\be
\Gamma_\mu \Gamma_\nu \ \ep = (\eta_{\mu\nu} -\vep_{\mu\nu\rho}\Gamma^\rho  )\ \ep
\ee
We find that
\be
[\delta_1,\delta_2] c^I_\mu = v^\nu  F^I_{\mu\nu} - \omega_{IJ} F^ J_{\nu\rho }\vep^{\nu\rho}_{\ \ \ \mu}
\label{bad}
\ee
with $\omega_{IJ}={i\over 2} \bar \ep _2 [\Gamma^I,\Gamma^J]\ep _1$.

The second term in (\ref{bad}) is a gauge transformation of the form
\[
c^I_\mu \to c^I_\mu + \epsilon_\mu {}^{\nu \lambda} \partial_\nu b^I_\lambda \; .
\]

We now consider the fermions. In light of the gauge symmetry (\ref{larger}), we need only verify that
\be
[\delta_1,\delta_2] \Gamma^\mu \eta_\mu = v^\nu \Gamma^\mu \partial_\nu \eta_\mu + \Gamma^\mu (\partial_\mu \chi)
\ee
for some $\chi$. For completeness, we should also check that the commutator of SUSYs acting on $\Psi_-$ includes a term
\[
[\delta_1,\delta_2] \Psi_- = \chi \; .
\]
Using the Fierz identity below, we find
\beas
[\delta_1,\delta_2] \Gamma^\mu \eta_\mu &=& {i \over 8} \Gamma^I \Gamma_\nu \Gamma^I \Gamma^\mu \Gamma^\alpha \partial_\mu \eta_\alpha \bar{\epsilon_2} \Gamma^\nu \epsilon_1 \cr
&&+{i \over 16} \Gamma^I \Gamma_{KJ} \Gamma^I \Gamma^\mu \Gamma^\alpha \partial_\mu \eta_\alpha \bar{\epsilon_2} \Gamma^{JK} \epsilon_1 \cr
&&+{i \over 384} \Gamma^I \Gamma_\nu \Gamma^{MLKJ} \Gamma^I \Gamma^\mu \Gamma^\alpha \partial_\mu \eta_\alpha \bar{\epsilon_2}\Gamma^{JKLM} \Gamma^\nu \epsilon_1 \cr
+ \Gamma^\mu (\partial_\mu \chi)
\eeas
Here, the last line vanishes in light of the relation
\[
\Gamma^I \Gamma^{JKLM} \Gamma^I = 0 \; .
\]
It is straightforward to show that the remaining terms take the form
\[
[\delta_1,\delta_2] \Gamma^\mu \eta_\mu = -2i \Gamma^\mu \partial_\nu \eta_\mu \bar{\epsilon_2} \Gamma^\nu \epsilon_1 + \Gamma^\mu \partial_\mu \chi
\]
for some $\chi$, as desired.

\subsection{Fierz Identity}
\setcounter{equation}{0}

The following relation is useful in verifying that the supersymmetry algebra closes:

For a general real symmetric $32 \times 32$ matrix $M$, we have
\[
M = {1 \over 32} \Gamma_{A} \tr(M \Gamma^A)+{1 \over 64} \Gamma_{BA} \tr(M \Gamma^{AB})+{1 \over 3840} \Gamma_{EDCBA} \tr(M \Gamma^{ABCDE}) \; .
\]
Applying this to
\[
M = \epsilon_2 \bar{\epsilon_1} - \epsilon_1 \bar{\epsilon_2}
\]
and using the fact that
\[
\Gamma^{012} \epsilon = \epsilon
\]
we find
\beas
\epsilon_2 \bar{\epsilon_1} - \epsilon_1 \bar{\epsilon_2} &=& ({1 \over 16} \Gamma_\mu + {1 \over 32} \Gamma_{\alpha \beta} \epsilon^{\alpha \beta} {}_\mu) \bar{\epsilon}_2 \Gamma^{\mu} \epsilon_1 \cr
&& + {1 \over 32} \Gamma_{JI}(1 + \Gamma^{012}) \bar{\epsilon}_2 \Gamma^{IJ} \epsilon_1 \cr
&& + {1 \over 384} \Gamma_\mu \Gamma_{LKJI} \bar{\epsilon}_2 \Gamma^{IJKL}  \Gamma^\mu \epsilon_1
\eeas
If we also have
\[
\Gamma^{012} \chi = \chi
\]
then
\beas
\epsilon_2 \bar{\epsilon_1}\chi - \epsilon_1 \bar{\epsilon_2}\chi &=& {1 \over 8} \Gamma_\mu \chi \; \bar{\epsilon}_2 \Gamma^{\mu} \epsilon_1 \cr
&& + {1 \over 16} \Gamma_{JI}\chi\; \bar{\epsilon}_2 \Gamma^{IJ} \epsilon_1 \cr
&& + {1 \over 384} \Gamma_\mu \Gamma_{LKJI}\chi \; \bar{\epsilon}_2 \Gamma^{IJKL}  \Gamma^\mu \epsilon_1
\eeas

\appendix{Gauge fixing the shift symmetry}

In this section, we discuss the proper gauge-fixing procedure starting with the classical action from section 3.1.

We focus on the bosonic gauge lagrangian
\be
{\cal L}_{\rm gauge} = (\aaa_\lambda  + \partial_\lambda X_-) \, \partial^\lambda X_+.
\ee
It has two gauge symmetries
\bea
\qquad &(1) & \qquad  \delta \aaa_\lambda = \partial_\lambda \alpha , \qquad  \delta X_- = \alpha \, ;  \\[2mm]
&(2) & \qquad \delta \aaa_\lambda =  \epsilon_{\lambda \mu \nu}  \partial^\mu \alpha^{\nu} \, .
\eea
The two symmetries allows imposing the two gauge conditions
\bea
\label{cond}
&(1)& \qquad
\partial^\nu  \aaa_{\nu} = 0, \qquad
\\[2mm]
& (2)& \qquad  \epsilon^{\lambda\mu\nu}\partial_\mu \aaa_\nu = 0\, .
\eea
The associated gauge fixing term (using a weighted gauge) and ghost action are
\be
% \epsilon^{\mu \nu \lambda} a_{\mu \nu} \, \partial_\lambda X_+
%{\cal L}_{\rm gauge} + {\cal L}_{\rm gf} + {\cal L}_{\rm ghost}
%\ee
{\cal L}_{\rm gf}  =\, \textstyle
\frac{1}{2} (\partial^\nu \aaa_\nu)^2 + \frac{1}{4}
(\epsilon^{\lambda\mu\nu}\partial_\mu \aaa_\nu)^2 =  \frac{1}{2} \partial^\mu \aaa^\nu \partial_\mu \aaa_\nu
%{\cal L}_{\rm ghost} \is
%\partial^\nu
\ee
\be
{\cal L}_{\rm ghost}  = \, \partial^\nu {\ccc}_+ \sp \partial_\nu \ccc_- \, -\frac{1}{2}\, (\partial_\mu\bar{C}_\nu - \partial_\nu \bar{C}_\mu)(\partial^\mu{C}^\nu - \partial^\nu {C}^\mu)
\ee
This ghost action introduces an additional gauge symmetry, so in this case, we need to add an additional gauge-fixing term for the $C$ ghost, and a ghost-for-ghost action\footnote{A possible choice
for this gauge fixing and ghost-for-ghost action is:
$
{1 \over 2}(\partial_\mu C^{I \mu})^2 -{1 \over 2} \partial_\mu g^I \partial^\mu g^I .
$}
but for now, let us omit this and leave the $C$ gauge-symmetry unfixed. The total action has a BRST symmetry:
\bea
\delta a_{\lambda} \iss \varepsilon(\partial_\lambda \ccc_- +  \epsilon_{\lambda\mu\nu}\partial^\mu C^{\nu}
% - \partial^\nu C^\mu)
)\,, \quad \qquad & \delta X^- = \, \varepsilon\sp \ccc_-\,, \\[2.5mm]
 \delta \ccc_+ \; = &  \varepsilon\sp \partial^\nu \aaa_\nu\,, \, \qquad \qquad  \qquad\qquad& \delta \bar{C}^\lambda =\, \sp \varepsilon \sp  \epsilon^{\lambda\mu\nu}\partial_\mu \aaa_\nu
 \eea
Now let us integrate out $\aaa_\nu$. The saddle point equation is
\be
 \aaa_\nu = \frac{1}{\square} \partial_\nu X_+.
 \ee
 The resulting action (after applying a shift $X_- \to X_- + \frac{1}{2 \square} X_+$) is simply
\be
\partial^\nu X_- \partial_\nu X_+\, +\, {\cal L}_{\rm ghost}
\ee
The BRST transformations simplify to
\be
\delta X_- = \varepsilon \sp \ccc_- \, ,  \qquad \qquad \delta {\ccc}_+ = \, \varepsilon\sp  \tilde{X}_+\, ,
\ee
where $\tilde{X}_+ \equiv \partial^\nu \frac{1}{\square}\partial_\nu X_+$ is the field $X_+$ with its constant zero mode removed. Thus, we end up with exactly the BRST-invariant action of section 3.2 plus the additional ghost action for $C^I_\mu$. But these vector ghosts are completely decoupled from the rest of the theory, both in the action and in the BRST transformations that define physical states. Thus, while the theory with a gauged shift symmetry technically still contains ghosts, it should be sensible to simply truncate the ghost sector of the theory and define the physical theory to be what remains. Equivalently, we can simply take the BRST-invariant action of section 3.2 (or the corresponding classical action of section 4.1) as our starting point.

\end{document}